\documentclass[twocolumn,prl,showpacs,preprintnumbers,amsmath,amssymb]{revtex4}

\usepackage{graphicx}
\usepackage{dcolumn}
\usepackage{bm}
\usepackage{bibtopic}

\begin{document}
%
\title{Massless Dirac fermions in epitaxial graphene on Fe(110)}

\author{A. Varykhalov, J. S\'anchez-Barriga, P. Hlawenka, O. Rader}

\address{Helmholtz-Zentrum Berlin f\"ur Materialien und Energie, 
Elektronenspeicherring BESSY II, Albert-Einstein-Stra\ss e 15, 12489 Berlin, Germany}

\begin{abstract}
Graphene   grown on Fe(110) 
by chemical vapor deposition using propylene   is investigated 
by means of angle-resolved photoemission. The presence of
massless Dirac fermions is clearly evidenced by the observation of a fully 
intact Dirac cone. Unlike Ni(111) and Co(0001), the Fe(110) imposes
a  strongly anisotropic quasi-one-dimensional 
structure on the graphene. Certain signatures
of a superlattice effect appear in the dispersion of its $\sigma$-bands but 
the Dirac cone does not reveal any detectable   
superlattice or quantum-size effects although the graphene corrugation is twice as large as
in the established two-dimensional graphene superlattice on Ir(111).

\end{abstract}


\maketitle

The interfaces of graphene with ferromagnetic transition metals and the resulting electronic strucuture
will be crucially important for the implementation of graphene in spintronics. 
Spin filter properties have been predicted for the epitaxial 
graphene/Ni(111) interface.\cite{Karpan-PRL-2007,Karpan-PRB-2008} 
Effective spin filtering requires hybridization between the exchange-split {\it 3d}-bands of
the substrate   and the $\pi$-bands of graphene.
This requirement as well as epitaxial growth are fulfilled in the graphene/Ni(111) and 
graphene/Co(0001) systems which have been widely studied in recent 
years both experimentally \cite{Varykhalov-PRL-2008,Varykhalov-PRB-2009,
Dedkov-PRL-2008} and theoretically. \cite{Karpan-PRL-2007,Karpan-PRB-2008,
Bertoni-PRB-2005,Voloshina-NJP-2011}

It has, however, been believed  that the characteristic linear dispersion
of Dirac fermions cannot exist in graphene/Ni(111) and graphene/Co(0001) due to
the exceptionally strong breaking of sublattice symmetry 
of graphene as a result of its chemisorption in the {\it top-fcc} geometry.
\cite{Oshima-SurfSci-1997,Heinz-NanoLett-2009}
In this case, the opening of a band gap at the Dirac point is expected. We have recently
examined the electronic structure of these systems by angle-resolved photoemission
(ARPES) 
and  discovered that the linear bands of Dirac fermions are neither
destroyed by electronic hybridization between graphene and Ni or Co
nor by the sublattice asymmetry of graphene. \cite{Varykhalov-PRX-2012} 
On the contrary, density-functional-theory study has shown that the effect of sublattice 
symmetry breaking becomes compensated by hybridization-induced 
charge transfer so that the Dirac cones are preserved. \cite{Varykhalov-PRX-2012}

While graphene on Ni(111) and Co(0001) have extensively been investigated, 
there are almost no reports about graphene on Fe surfaces, despite the larger
magnetic moment of Fe. 
More detailed studies could also become important in the context of extrinsic carbon magnetism. 
The lack of investigations is possibly
related to difficulties of graphene preparation on Fe. Indeed, our experience 
shows that chemical vapor deposition (CVD) of hydrocarbons at the parameters 
typical for Co and Ni substrates leads to the formation of iron surface carbide instead
 of graphene. A rare successful example has been the creation of a 
graphene/Fe(111) interface by intercalation of Fe between graphene 
and Ni(111). \cite{Dedkov-APL-2008,DedkovPCCP11}  The intercalated 
Fe films are ultrathin (a few monoatomic layers) and ARPES spectra have been reported in normal emission ($\overline{\Gamma}$).\cite{Dedkov-APL-2008,DedkovPCCP11}

A breakthrough in the fabrication of graphene on Fe surfaces has recently been reported.
Vinogradov et al. \cite{Vinogradov-PRL-2012} found that the formation of iron carbide can be avoided in favor of 
graphene growth by significantly rising the partial pressure of carbon-containing molecules during 
the CVD procedure. Ref. \onlinecite{Vinogradov-PRL-2012} has
demonstrated the formation of a high-quality graphene layer on Fe(110) 
with pronounced quasi-1D moir\'e pattern emerging from
different lattice symmetries of graphene and the (110)-surface of bcc Fe.
Graphene/Fe(110) has extensively been characterized by low-energy electron diffraction (LEED), 
scanning tunneling microscopy, core-level photoemission, and x-ray absorption spectroscopy
but not by measurements of the band structure.\cite{Vinogradov-PRL-2012}
The band dispersions in 1D-modulated
layers with a lateral periodicity of a few nanometers can be   subject 
to lateral quantization \cite{Ortega-SS-Vicinal-1,Ortega-SS-Vicinal-2,
Varykhalov-Au-Stripes-PRB-2005} which is particularly interesting in the case of 
graphene.\cite{Katsnelson06,Cheianov06,Ando98}

In the present paper, we study the band structure of graphene synthesized on Fe(110) in detail. 
The experimental data have been 
acquired by means of ARPES. Neither overview measurments of the 
electronic structure of graphene/Fe nor high-resolution data
of the $\pi$-band in the vicinity of $\overline{K}$-point of the Brillouin zone of 
graphene,
reveal any significant differences to the cases of graphene on Ni(111) and 
Co(0001).\cite{Varykhalov-PRX-2012} The ARPES measurements demonstrate 
a fully intact Dirac cone at the $\overline{K}$-point also in graphene on Fe(110).
The $\sigma$-bands reveal   faint replicas which indicate
 an {\it umklapp} process due to 1D-modulation of graphene/Fe(110) but
no significant quantum-size effects are revealed in the 
dispersion of the $\pi$-band.

\begin{figure}[t]
\centering
\includegraphics[width=0.42\textwidth]{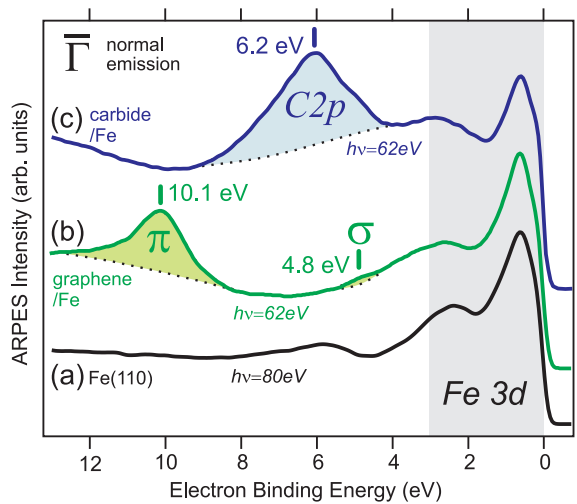}
\caption{
Normal emission valence band spectra of (a) bare Fe(110) grown on W(110),
(b) Fe(110) with graphene, (c) Fe(110) with surface carbide. 
}
\end{figure}

\begin{figure}[t]
\centering
\includegraphics[width=0.48\textwidth]{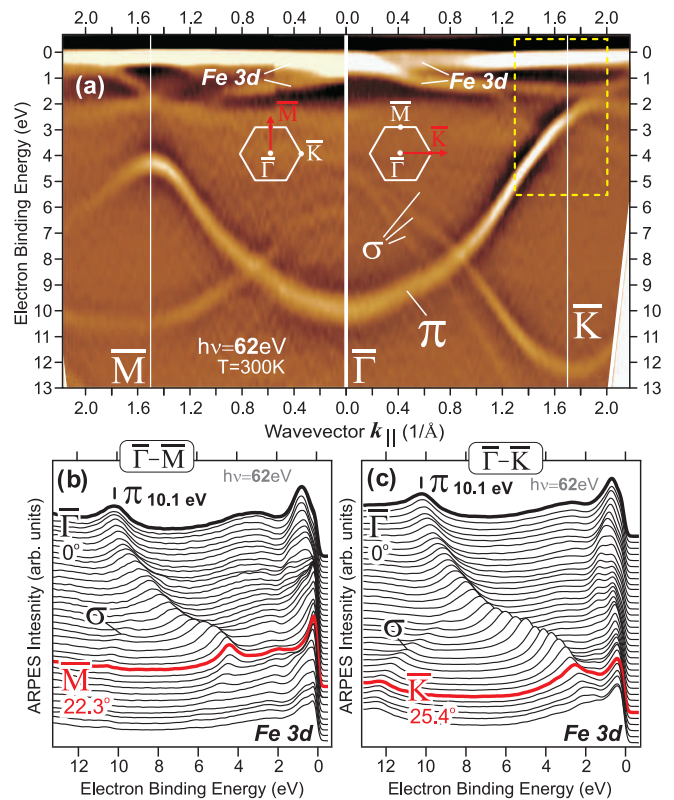}
\caption{
(a) Overall band structure of graphene on Fe(110) sampled by ARPES along
$\overline{\Gamma}-\overline{K}$ and $\overline{\Gamma}-\overline{M}$ 
directions of the surface Brillouin zone in a color-scale representation. 
(b) Dispersion along $\overline{\Gamma}-\overline{M}$ as ARPES spectra.
(c) Dispersion along $\overline{\Gamma}-\overline{K}$ as ARPES spectra.
}
\end{figure}

\begin{figure}[t]
\centering
\includegraphics[width=0.48\textwidth]{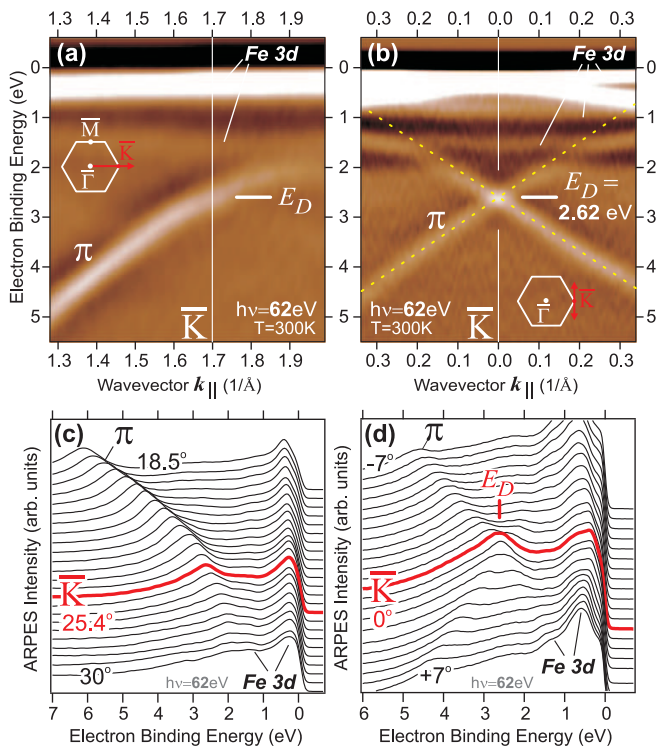}
\caption{Characterization of graphene/Fe(110) by high-resolution ARPES.
(a) Dispersion of the $\pi$-band near the $\overline{K}$-point along 
$\overline{\Gamma}-\overline{K}$.
(b) Intact Dirac cone revealed in the dispersion of the $\pi$-band in the $\perp$$\overline{\Gamma}-\overline{K}$
direction. ARPES data in (a) and (b) are presented as second 
derivative over the energy. Raw data corresponding to (a) and (b)  are displayed as 
spectra in (c) and (d), respectively.
}
\end{figure}

\begin{figure}[t]
\centering
\includegraphics[width=0.45\textwidth]{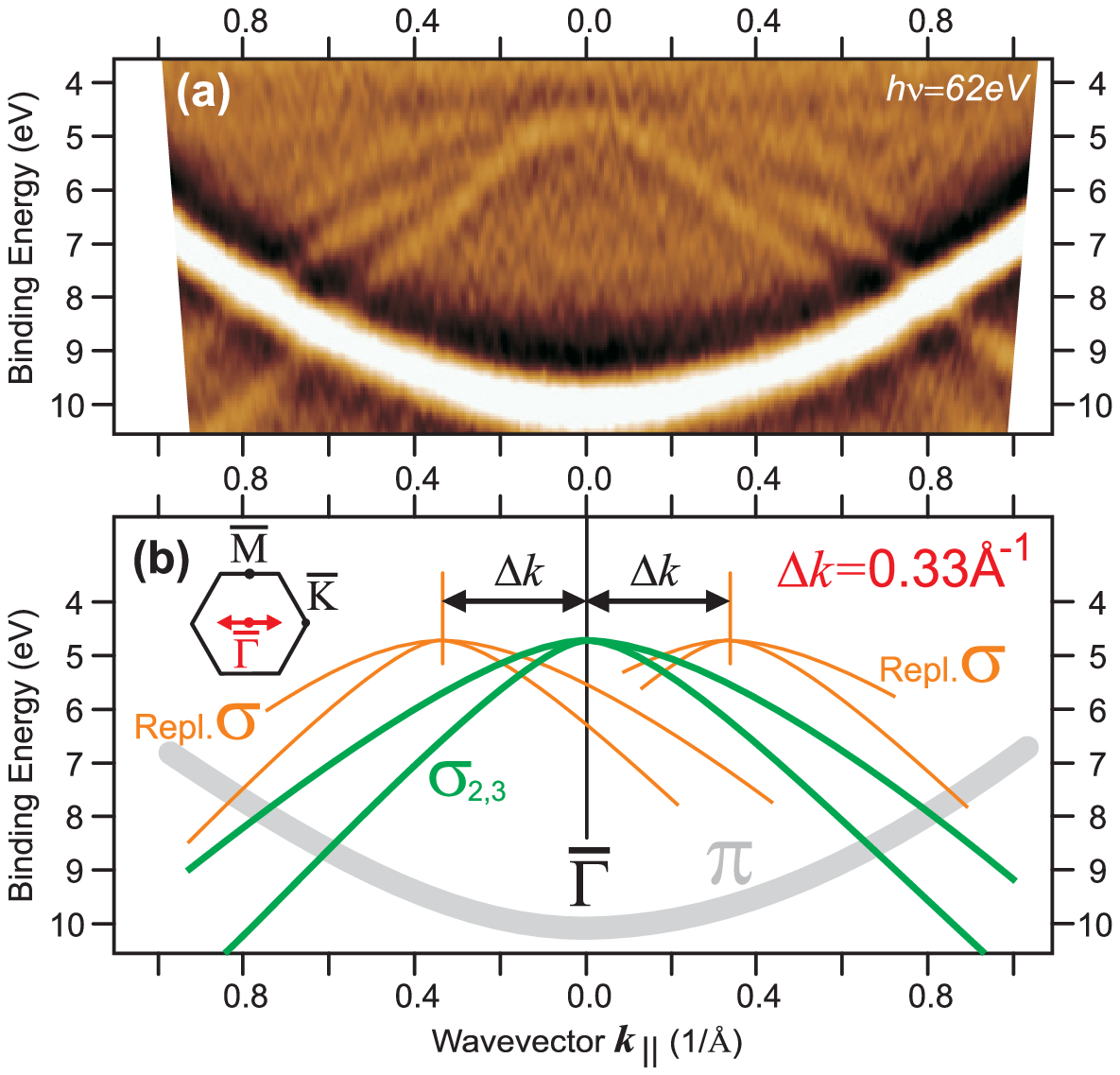}
\caption{
(a) High-resolution measurements of the $\sigma$-band dispersion near the
$\overline{\Gamma}$-point. The second derivative over the energy is shown.
Faint replicas displaced by $\Delta${\it k}$\sim\pm$0.33\AA$^{-1}$ from $\overline{\Gamma}$  
are seen. (b) Band structure of replicas extracted from ARPES data shown in (a).
}
\end{figure}

The Fe(110) has been prepared as 40-monolayer film of Fe deposited and annealed on top of W(110).
Graphene was fabricated by CVD of propylene as described in Ref. \onlinecite{Vinogradov-PRL-2012}.
The structural quality of the Fe(110) layer and of the synthesized 1D-modulated graphene were tested by LEED 
and photoemission. ARPES measurements have been conducted at the UE112-PGM2a beamline of BESSY II
using the end station ARPES 1$^{2}$. A photon energy of h$\nu$=62 eV and linear (mixed s+p)
light polarization have been  used. Measurements have been  performed at room temperature.
Figure 1 shows valence band spectra of Fe(110) before [Fig. 1(a)] and after [Fig. 1(b)]
synthesis of the graphene overlayer. Spectra were measured at the $\overline{\Gamma}$-point
of the Brillouin zone ({\it k}$_\parallel$=0 \AA$^{-1}$). The formation of graphene
is evidenced by an intense photoemission peak of the $\pi$-band at 10.1 eV binding energy as well as a small peak
corresponding to $\sigma$-bands at 4.8 eV. It is worthwhile noting  that the binding energy of
the $\pi$-band at $\overline{\Gamma}$ is very close to that for graphene on Ni(111) and Co(0001) (Ref. 
\onlinecite{Varykhalov-PRX-2012}) indicating  that the chemical interaction of graphene 
with Fe is strong, exactly as with Ni and Co. For the purpose of comparison, Fig. 1(c) shows
a valence band spectrum of fully carbidized Fe(110) which is typically obtained when CVD is performed 
at a low partial pressure of hydrocarbons. The iron surface carbide is identified by a very broad peak at
$\sim$6.2 eV which does not disperse with the emission angle (wavevector {\it k}$_\parallel$).

Figure 2(a) displays in color scale the overall band structure of graphene on Fe(110) sampled along
$\overline{\Gamma}-\overline{K}$ and $\overline{\Gamma}-\overline{M}$ directions of
the graphene Brillouin zone (sketched in the insets). Figures 2(b) and 2(c)
additionally emphasize the dispersions by presenting ARPES spectra for different emission angles.
One sees that the structural quality of the graphene is high, as no significant contribution from
rotational domains occurs (such appears for graphene/Co(0001) \cite{Varykhalov-PRX-2012}.).
Besides   faint additional bands of $\sigma$-type near the $\overline{\Gamma}$-point
(see further below) the band structure is almost identical to 
that of graphene on Ni and Co. \cite{Varykhalov-PRX-2012}
It is also clearly seen
that the $\pi$-band disperses through the $\overline{K}$-point of the Brillouin zone without any significant
distortion (see Figs. 3(a) and 3(c) which zoom the area marked in Fig. 2(a) by a yellow dashed frame) .

To investigate the behavior of the $\pi$-band at the $\overline{K}$-point more in detail, we have
sampled its dispersion along the direction $\perp$$\overline{\Gamma}-\overline{K}$,
where both sides of the conical dispersion are seen. \cite{Varykhalov-PRX-2012,Himpsel-BZ-Effects}
The result is displayed in Fig. 3(b) and emphasized through the second derivative of the photoemission
intensity over the energy.
Raw ARPES data are displayed in Fig. 3(d) as stacked spectra. 
One clearly sees an intact Dirac cone with intense Dirac crossing point. The Dirac point is  
located at at 2.62 eV binding energy which manifests a significant {\it n}-doping of graphene.
The experimental dispersion near the $\overline{K}$-point perfectly fits with the linear bands 
of ideal graphene obtained from the tight-binding model \cite{Wallace-PRB-1947} [yellow dashed lines in Fig. 3(b)] 
after a corresponding energy shift.
The observation of  the intact Dirac cone clearly indicates the presence of massless Dirac fermions 
in graphene on Fe(110).
Similarly to the case of graphene on Ni and Co, \cite{Varykhalov-PRX-2012} the folding of the
$\pi$-band at lower binding energy due to hybridization with Fe {\it 3d}-bands 
and occurrence of a hybridization gap are observed. 

The direction $\perp$$\overline{\Gamma}-\overline{K}$ along which
the dispersions from Fig. 3(b) has been measued corresponds to 
electrons propagating perpendicularly to the quasi-1D stripes of the graphene moir\'e on Fe(110).
It is interesting that the Dirac cone, having proven sensitive to superlattice effects,
\cite{minigaps-ARPES-Ir111,Sanchez-Barriga-Clusters} exhibits no features which can be associated 
with any effects of superlattice of lateral quantization (e.g. band replicas or energy gaps). This is rather surprising 
since the amplitude of the structural corrugation of the 1D graphene moir\'e
determined as 0.6-0.9\AA\ by scanning tunneling microscopy and density-functional theory
 in Ref. \onlinecite{Vinogradov-PRL-2012} 
is at least two times larger than that of graphene/Ir(111) for which moir\'e-induced quantum-size
effects are very pronounced. \cite{minigaps-ARPES-Ir111,Sanchez-Barriga-Clusters,
Starodub-PRB-2011}  

The only signature of quantum-size effects was found in the
$\sigma_{2,3}$-band of graphene near the center of the SBZ
and along $\overline{\Gamma}-\overline{K}$ direction.  The effect is 
weak and can only be observed in the second derivative of the ARPES signal. 
Measured dispersions of $\sigma_{2,3}$-bands are zoomed in Fig. 4(a),
Fig. 4(b) displays the extracted band structure. Faint {\it umklapp}-like
replicas displaced by $\Delta${\it k}=$\pm$0.33\AA$^{-1}$ from 
$\overline{\Gamma}$ are seen. The real-space periodicity {\it D}=$\frac{2\pi}{\Delta k}$ 
corresponding to $\Delta${\it k} is about 16\AA, which differs by far from
the periodicity of quasi-1D graphene stripes reported in Ref. \onlinecite{Vinogradov-PRL-2012} ($\sim$40\AA).
However, the direction $\overline{\Gamma}-\overline{K}$ along which the dispersion
in Fig. 4(a) was measured is the direction {\it along} the stripes, and, in principle, 
should not demonstrate any quantization effects. Ref. \onlinecite{Vinogradov-PRL-2012}
has shown that the nanostripes of the  graphene moir\'e on Fe(110) 
also possess a complex longitudinal wave-like pattern with a periodicity of $\sim$19\AA.
This value fits much better with the extracted $D$=16\AA\   and suggests that the 
observed replication of $\sigma$-bands is {\it umklapp} emerging
along the stripes due to the wave-type pattern. The question  why no
{\it umklapp}-induced replicas can be observed for the Dirac bands from
Figs. 3(a) and 3(b) 
can be a subject for theoretical studies, since it has been shown that  $\pi$- and $\sigma$-states
must be equally sensitive to external superpotentials and 
superstructural corrugations. \cite{Usachov-PRB-2012}

In summary, we have grown graphene on Fe(110) according to the method
proposed by Vinogradov et al. \cite{Vinogradov-PRL-2012} and characterized its
electronic structure by angle-resolved photoemission. The obtained results are very 
similar to the band structures of graphene on Ni(111) and Co(0001) established in
our earlier investigation. \cite{Varykhalov-PRX-2012}
Near the $\overline{K}$-point of the surface Brillouin zone 
the $\pi$-band of graphene on Fe(110) reveals a significant {\it n}-doping, 
a fully intact Dirac cone and pronounced electronic hybridization with the {\it 3d}-bands 
of the ferromagnetic Fe substrate. Surprisingly, we did not observe any measurable 
quantum-size effects in the conical dispersion of Dirac fermions despite the
quasi-1D moir\'e pattern of graphene on Fe(110) 
and weak signatures of {\it umklapp} in its $\sigma$-bands.

Acknowledgments. This work was supported by SPP 1459 of the Deutsche
Forschungsgemeinschaft.

\end{document}